# Fat Tails in Financial Return Distributions Revisited: Evidence from the Korean Stock Market

Cheoljun **Eom**[a], Taisei **Kaizoji**[b], and Enrico **Scalas**[c]

a. School of Business, Pusan National University, Busan 46241, Korea
b. Graduate School of Arts and Sciences, International Christian University, Tokyo 181-8585, Japan
c. Department of Mathematics, School of Mathematical and Physical Sciences, University of Sussex, Brighton, BN1 9QH, United Kingdom

**ABSTRACT**

This study empirically re-examines fat tails in stock return distributions by applying statistical methods to an extensive dataset taken from the Korean stock market. The tails of the return distributions are shown to be much fatter in recent periods than in past periods and much fatter for small-capitalization stocks than for large-capitalization stocks. After controlling for the 1997 Korean foreign currency crisis and using the GARCH filter models to control for volatility clustering in the returns, the fat tails in the distribution of residuals are found to persist. We show that market crashes and volatility clustering may not sufficiently account for the existence of fat tails in return distributions. These findings are robust regardless of period or type of stock group.

# 1. INTRODUCTION

Since the 1950s, for financial markets, empirical stylized facts such as volatility clustering, price reversals, asymmetric distributions, fat tails, and time-varying properties that are commonly observed in financial time series have originated new financial theory and model expansion (see e.g., Cont [1]). Market crashes affect these stylized facts, particularly the existence of fat tails in return distributions. As a result, the topic of fat tails in return distributions has attracted substantial academic and practical attention aimed at improving risk management. Volatility has been identified as a crucial factor for risk measurement in finance. The property of short-term and long-term persistence in volatility is deeply related to the volatility clustering usually observed in financial time series. The GARCH model proposed by Bollerslev [2-3] is an extremely useful tool for describing the influence of volatility clustering and its relationship to fat-tails in return distributions.[1] Using the daily returns of the U.S. Dow Jones Industrial Average index, Rachev et al. [11] presented evidence that the fatness of the tails in return distributions tends to increase during market crashes. Stiyanov et al. [12] presented evidence that the existence of fat tails in return distributions cannot be fully described by volatility clustering because fat tails are still observed in the distribution of standardized residuals estimated from the GARCH model. Previous studies have considered market crashes and volatility clustering as important factors affecting the existence of fat tails in return distributions.

Among the earliest studies of the distribution of stock returns are those of Mandelbrot [13], Fama [14] and Samuelson [15] (see also Cootner [16] for an early review). They reported that the empirical distribution of a return series has properties different from those of the normal distribution, with a more peaked central section and much fatter tails, and suggested that these properties were closer to the properties of an α-stable distribution. However, even though the α-stable distribution has fatter tails than the normal, the infinite variance and constant fatness of the tails in the distribution seem to differ from empirical properties observed in the distribution of a return series. Cont [1] presented negative evidence for the existence of infinite variance based on an investigation of the idea that the variance of the stock return converges to a certain level as sample size increases. Mantegna and Stanley [17] and Ponta et al. [18] presented evidence that the fatness of the tails in return distributions is not constant, in accordance with changes in the measurement time interval used to convert price data into return data. Given these arguments, researchers have looked at Student's $t$ distribution as another possible representative distribution having the property of fat tails. Compared to the tails of a normal distribution, the tails in Student's $t$ distribution can be made fatter by adjusting the degrees of freedom parameter. Praetz [19] and Blattberg and Gonedes [20] showed that Student's $t$ distribution has similar distributional properties to those observed for actual returns. Peiro [21] presented evidence that Student's $t$ distribution in stock markets such as those in the United States, Japan, United Kingdom, Germany and France is very close to the empirical distribution of returns. Zumbach [22] showed the usefulness of the risk estimation model based on Student's $t$ distribution with five degrees of freedom, using the return data of FTSE 100. However, even though Student's $t$ distribution has fatter tails than the normal, this distribution fails to describe the properties of a more peaked central section and the asymmetric structure of the distribution observed in return series. Moreover, it is difficult to sufficiently describe the dynamic and distributional properties of returns using a fixed value for the degrees of freedom estimated in Student's $t$ distribution because of the time-varying property of returns, whereby the property is continuously changing according to the state of the market. Also, due to the absence of a universally accepted theory for the "true" distribution of returns, many of these research efforts have focused on empirical investigations using statistical methods. Both the α-stable distribution and Student's $t$ distribution are still used to describe return distributions in the field of finance because they are convenient for analytical and numerical calculations.[2]

---

[1] In finance, accurate measurement and good forecasting of volatility play a crucial role as risk factors in asset pricing, optimal portfolio allocation, and option pricing models. Standard models related to volatility are typically as follows: ARCH [4] and GARCH [2-3] models, implied volatility from the Black-Scholes option pricing model [5], and volatility indices from squared or absolute values of returns. In addition, a model-free approach of realized volatility using high-frequency intraday returns was proposed by Andersen and Bollerslev [6] and previous studies [7-10] applied it in finance. Based on our research goals, the present study employs the standard models of the GARCH family to reflect the characteristics of volatility clustering and fat tail among stylized facts in the financial time series.

Our study investigates the existence of fat tails in the return distributions based on appropriate statistical methods. Specifically, we aim to verify the existence of fat tails in the return distribution by conducting a comparative investigation between theoretical and empirical return distributions. We control for the effects of market crash and volatility clustering by classifying sub-periods and using the GARCH filter models. The purpose here is to determine whether market crashes and volatility clustering are sufficient to explain the existence of fat tails in return distributions. In this study, the fatness of the tails is quantified by the *statistical probability* defined as a measurement of the empirical probability of the frequency located in the tails, where the number of return data points belonging to each tail area—that is, the areas outside the 99% central section of the distribution—is divided by the total number of return data points in the distribution. Our results are summarised briefly as follows. We verify the existence of fat tails in return distributions for the sub-periods both before and after the 1997 Korean foreign currency crisis, which we use as a representative market crash in the Korean stock market, and for all the various stock groups classified by market capitalization. Furthermore, the size of the fat tails in the more recent sub-periods is much greater than in the pre-crash sub-periods, and the return distribution extracted from the small-capitalization stocks (small-cap) has a fatter tail than the tail for the large-capitalization stocks (large-cap). After controlling for the 1997 market crash and for the volatility clustering phenomenon in the return series by using the GARCH filter models, the fat tails in the distribution remain. These results are robust when using other GARCH models such as the best-fit ARMA-GARCH model and the best-fit EGARCH model. Therefore, we conclude that market crashes and volatility clustering may not fully account for the existence of fat tails in return distributions. As a result, neither market crashes nor volatility clustering explains the fat tails completely. Accordingly, based on our findings, we expect future studies to uncover the economic meaning of the risk property included in the fat tails of return distributions, thereby serving as a bridge connecting existing studies (e.g., Kelly and Jiang [26]) of tail risk in pricing models.

---

[2] In finance, there are other types of distributions that allow for properties of leptokurtosis and skewness. For example, the use of time-change leads to a finite-variance subordinated stochastic process [23], and the mixture distributions assume that stock returns have a mixed distribution under the interaction of a continuous diffusion and a discontinuous jump process [24-25]. This paper uses three types of distribution, i.e., normal, Student's t and α-stable, to study the properties of fat tails and higher central peaks in the return distribution.

The paper is organised as follows. Section 2 describes the data, the periods of observation and the methods used for testing the research hypotheses. Section 3 presents the results on the existence of fat tails in return distributions. Section 4 provides the summary and conclusions.

## 2. EMPIRICAL DESIGN

### 2.1. Data and Periods

In this study, we use the daily logarithmic returns or log-returns ($r_t = \ln P_t - ln P_{t-1}$, where $P_t$ is the price at day $t$) of the market and individual stocks traded on the Korean stock market from January 1980 to June 2015 (sample size $T$=8,719).[3] This full period is used to compare the sub-period of the 1997 Korean foreign currency market crash with sub-periods of normal markets, under the assumption that the event of market crash is a factor influencing the fat tails in return distributions. The 1997 Korean foreign currency crisis was chosen as a representative market crash in the Korean stock market. Following the 1997 Asian financial crisis, the Korean government received formal financial support from, and underwent structural adjustment imposed by, the International Monetary Fund for roughly three years, beginning in December 1997. For this study, we chose two types of sub-period. The first consists of two sub-periods of normal market activity before and after the market crash: the former sub-period is from July 1982 to June 1997 and the latter sub-period from July 2000 to June 2015. The second set of sub-periods includes three sub-periods: a sub-period for the market crash and two sub-periods of the normal market. Specifically, we defined a prior-to-crash sub-period from July 1988 to June 1997, a market crash sub-period from July 1997 to June 2006, and a more recent sub-period from July 2006 to June 2015. In addition, we identified three stock groups classified by the market capitalization (= price×number of outstanding shares) of firms. Since the study of Banz [27], firm size has been one of the key factors to explain changes in stock returns; accordingly, we considered three groups: an all-stock group, a large-cap group (consisting of the stocks of firms whose market capitalization ranked them in the top 40% among all firms), and a small-cap group (stocks of firms whose capitalization placed them in the bottom 40% among all firms). The classification of stock groups is based on the average value of monthly firm market capitalization within a particular period. In addition, **Table 1** reports the summary statistics of the stock returns mentioned above.

---

[3] The data source is FnGuide: a data vendor in Korea.

[insert **Table 1** in here]

The table shows descriptive statistics of average value, standard deviation, skewness, and kurtosis of stock returns. As is well known, the period (1997.07~2006.06), which includes the event of market crash, has a lower average value and higher standard deviation than the normal market periods. The high-order moments of skewness and kurtosis deviate from those expected for the normal distribution. Interestingly, large-cap stocks have higher average value and lower standard deviation of returns, compared to small-cap stocks. This contrasts with the size effect [27], in which the small-cap stocks have higher standard deviation and higher average value of returns than large-cap stocks. Based on a previous study [28], this suggests that the 1997 foreign currency crisis in the Korean stock market may have led to a change to the trading environment centred on large-cap stocks. Further investigation into this phenomenon is beyond the scope of the present study and forms an avenue for future research.

## 2.2. Methods

This section describes the main features of the methods used for investigating the existence of fat tails in return distributions. In applying these methods, we controlled for the influence of market crashes and volatility clustering and compared various theoretical distributions to the corresponding empirical distributions.

The baseline model in financial theory assumes that log-returns are normally distributed so that prices follow the log-normal distribution. This is used in the so-called Black-Scholes-Merton option pricing model [5]. However, the empirical distribution observed from actual returns data has a more peaked central section and much fatter tails than the normal distribution, making it a leptokurtic distribution. As mentioned in the Introduction, as alternatives to the normal distribution, other theoretical distributions that might better describe such an empirical behaviour include the α-stable distribution and Student's $t$ distribution. The probability density function of a Student $t$ random variable $X$ [29] is defined as follows:

$$f(x|\nu) = \frac{\Gamma\left(\frac{\nu+1}{2}\right)}{\Gamma\left(\frac{\nu}{2}\right)} \frac{1}{\sqrt{\nu\pi}} \frac{1}{\left(1+\frac{x^2}{\nu}\right)^{\frac{\nu+1}{2}}} \quad (1)$$

where $\nu$ represents the degrees of freedom, and $\Gamma(\cdot)$ is the Gamma function. The stable distribution, $S(\alpha, \beta, \gamma, \delta; 1)$, is characterized by four parameters. There is no general analytic form for the probability density function of a stable-distributed random variable, $X$, but its characteristic function can be written as follows [30-31]:

$$\hat{f}(\kappa|\alpha,\beta,\gamma,\delta) = \mathbf{E}[\exp(i\kappa X)] = \begin{cases} \exp\left(-\gamma^\alpha |\kappa|^\alpha \left[1 - i\beta\left(\tan\frac{\pi\alpha}{2}\right)(\text{sign } \kappa)\right] + i\delta\kappa\right), & \alpha \neq 1 \\ \exp\left(-\gamma|\kappa| \left[1 + i\beta\frac{2}{\pi}(\text{sign } \kappa)\log(|\kappa|)\right] + i\delta\kappa\right), & \alpha = 1 \end{cases} \quad (2)$$

where, $\alpha$ $(0 < \alpha \leq 2)$ is the tail index, $\beta$ $(-1 \leq \beta \leq 1)$ denotes the skewness of the distribution, and $\gamma$ $(0 < \gamma < \infty)$ and $\delta$ $(-\infty < \delta < \infty)$ are the scale and the location parameters, respectively. Several different parametrizations of the characteristic function are possible and we use the convention set by Nolan [31] for the S1 parametrization used by Veillette. We use three distributions, Student's t, α-stable, and normal, and compare them with the empirical distribution observed from return data. The test process is as follows. We first estimate the parameters of each theoretical distribution from the actual return data.[4] Using the estimated parameters, we generate random numbers. The frequency distribution

---

[4] We employ the MATLAB tool to estimate parameters of each theoretical distribution from the actual return data. The parameters of the α-stable distribution are fitted by the ordinary least square regression, using the *STBL* m-code provided by M. Veilette based on Koutrouvelis [32-33]. The parameters of the normal and Student's t

extracted from the random numbers is then compared to the empirical distribution from the actual returns. The comparison between theoretical and empirical distributions is divided on the central sections and the tails. The central section in the distribution is compared by means of a probability density function plot (a normalized histogram); the tails are compared by using a double-logarithmic plot based on the cumulative distribution function, as in Mantegna and Stanley [17]. The bin size of the frequency distribution is determined according to Scott [34].

The following logic and method were used to investigate the effect of market crashes on the fat tails of return distributions. The tails of return distributions represent large losses and large profits that deviate from the average value. Thus, a market crash, which causes large-scale price fluctuations, ends up in the tails of a return distribution. If during a sub-period of a market crash we measure a return distribution with much fatter tails compared to the sub-periods of normal market, this is evidence suggesting that market crashes are a significant factor influencing the tails of return distributions. In our test, we divide the entire period of observation into before- and after-crash sub-periods. We standardize the return data for all periods by subtracting the average value and dividing by the standard deviation. We define the tails of distribution as the extreme areas of the empirical frequency distribution for the standardized returns at either tail 0.5% area outside of the central 99% area of the standard normal distribution. In other words, for calculating the fatness of tails, we use the frequency of the standardized returns that are smaller than -2.58 and greater than 2.58. This *statistical probability* is given by the relative frequency ($f_N/f_T$) ratio, calculated as the number of values ($f_N$) included in the tails of the distribution divided by the total number of values ($f_T$). We then use this statistical probability to characterize the fatness of the tails. A statistical probability greater than 0.005 is evidence supporting the existence of fat tails.

Finally, we suggest a method for controlling the effect of volatility clustering on the fat tails in return distributions by using the GARCH model as a filter model. Volatility clustering is strongly related to the persistence of volatility, wherein large-scale price fluctuation tends to be followed by large-scale price fluctuation. The fat tails in return distributions have a close relationship to volatility clustering since all large-scale price fluctuations are located in the tails of return distributions. The GARCH model of Bollerslev [2-3] is a useful tool that effectively reflects the relationship between volatility clustering and the fat tails of return distributions. Hence, we empirically assess the effect of volatility clustering on the fat tails through the GARCH (1,1) filter model. In the GARCH (1,1) model, we have for the log-

---

distributions are fitted using *fitdist m*-code in the statistics toolbox of MATLAB.

returns $r_{j,t}$:

$$r_{j,t} = \mu_j + \varepsilon_{j,t} \tag{3a}$$

$$\sigma_{j,t}^2 = d + \beta_j \sigma_{j,t-1}^2 + \gamma_j \varepsilon_{j,t-1}^2 \tag{3b}$$

where the index $j$ represents the $j$-th stock, $d > 0$, $\beta_j > 0$, $\gamma_j > 0$, and $\beta_j + \gamma_j < 1$, so that the variance of stock returns, $\sigma_{j,t}^2$, is a linear combination of the last-period variance, $\sigma_{j,t-1}^2$, and the last-period squared residuals, $\varepsilon_{j,t-1}^2$, and $\varepsilon_{j,t}$ are random variables following the normal distribution with variance $\sigma_{j,t}^2$. If the statistical probability associated with the tail-end values after controlling for volatility clustering is significantly reduced in comparison to results using the actual return data, this is evidence suggesting that volatility clustering is a significant factor influencing the existence of fat tails in return distributions. The test process is as follows. We estimate the standardized residuals ($\varepsilon_{j,t}^*$) using the GARCH filter model for the return data in each period. The standardized residual is a residual ($\varepsilon_{j,t} = r_{j,t} - \mu_j$) divided by the conditional standard deviation, that is, $\varepsilon_{j,t}^* = \varepsilon_{j,t}/\sigma_{j,t}$. As before, the statistical probability using the estimated standardized residuals is then calculated to quantify the fatness of the tails of their distribution.

## 3. RESULTS

As previously mentioned, we use three theoretical distributions—normal, α-stable and Student's $t$—and compare each one to the empirical distribution using the market return data. **Figure 1** shows results from the comparison of actual market returns to each of the three theoretical distributions. The market returns are for KOSPI, a representative index of the Korean stock market, over the period from January 1980 to June 2015 ($T$=8,719). The three theoretical distributions in the figure are histograms of random number data generated using distribution parameters directly estimated from the market returns. Specifically, we use the standardized data of the market returns and the random number data to define each frequency distribution. The figure separately shows the central section of the four distributions (**Figure 1(a)**) and the tails of the four distributions (**Figure 1(b)** and **Figure 1(c)**). **Figure 1(a)** shows the central section of the distributions using the probability density function; **Figures 1(b)** and **1(c)** present the negative and positive tails of the distributions using a double-logarithmic plot based on the cumulative distribution function, respectively.

[Insert **Figure 1** in here]

From **Figure 1**, we can confirm visually that the empirical distribution of market returns has fat tails. While the central section of the empirical distribution is not far from the theoretical distributions, the tails in the empirical distribution show a clear difference. The descriptive statistics calculated from the market returns are a mean of 0.000339, standard deviation of 0.0162, skewness of -0.1388, and kurtosis of 8.5081. The estimated parameters of each of the three types of theoretical distribution are reported in **Table 2**. In **Figure 1(b)** and **1(c)**, the tails in the empirical distribution of market returns are much fatter than in the normal distribution but thinner than in both the α-stable distribution and Student's $t$ distribution. For robustness, we use the Kolmogrov-Smirnov (KS) test as a quantitative estimate of the distance between each one of the theoretical distributions and the empirical distribution in **Figure 1**. Results are reported in **Table 2**. Based on the empirical distribution, the null hypothesis is that the two samples are drawn from the same underlying distribution. We employ the simulation that is repeatedly performed using each 100-random number dataset generated using parameters of each distribution directly estimated from the market returns. All KS statistics for theoretical distributions present evidence that significantly rejects the null hypothesis, compared to the empirical distribution.

[Insert **Table 2** in here]

In addition, **Figure 2** shows that there is a comparative point of statistical probability on the central and tail parts in the empirical distribution based on the central 95% area of the standard normal distribution. That is, based on the choice of probability of 0.90 for the central part and 0.05 of the negative and positive tail parts in the standard normal distribution, the statistical probability of the return distribution shows a higher value of 0.9184 and lower values of 0.0418 and 0.0396, respectively. On the other hand, compared to the 0.99 of the central part and the 0.005 of the negative and positive tail parts in the standard normal distribution, the statistical probability of the return distribution shows a lower value of 0.9709 and higher values of 0.0146 and 0.0143, respectively. According to the empirical evidence of **Figure 2**, using the probability of 0.005 in the tails of the return distribution is justified as a criterion to evaluate the existence of fat tails.

[Insert **Figure 2** in here]

Next, further results regarding the fatness of the tails in the empirical distributions of stock returns are presented in **Table 3**. The fatness of the tails is measured by the statistical probability and the estimated parameters of Student's $t$ distribution, that is, the degrees of freedom. In order to control for the effect of extreme values on the results, we calculated statistical probabilities and estimated parameters after removing the top and bottom 5% of stocks based on value among all stocks selected in each period. Panel A in the table presents the average values of the statistical probabilities and Panel B presents the average values of the estimated parameters of Student's t distribution. The results are divided according to period and stock group, i.e., the entire period and the sub-periods before and after the market crash, and the three stock group classifications—all stocks, large-cap and small-cap—as determined by the market capitalization of the firms involved.

[Insert **Table 3** in here]

**Table 3** shows that regardless of period and stock group, the distributions of the stock returns have fat tails. All statistical probabilities in Panel A are greater than 0.005, and all estimated parameters of Student's t distribution in Panel B are very small compared to the normal distribution limit of Student's *t* (more than 30 degrees of freedom). In terms of period, regardless of stock group, the fatness of tails in the stock return distributions is on average greater in the recent sub-periods than in the earlier sub-periods. The results from the estimated parameters of Student's t distribution are similar with the results of Panel A. These results suggest that the recent sub-periods in the Korean stock market have more frequent large-scale price fluctuations compared to the earlier sub-periods. A possible reason is that the Korean government transitioned to a capital liberalization policy for foreign investors after the 1997 Korean foreign currency crisis; hence the recent periods have a market more open to foreign investors as compared to the earlier periods. This finding is consistent with Boyer et al. [35], who found that a financial market opens to foreign investors experienced a greater impact from the 1997 Asian foreign currency crisis. In addition, the fatness of the tails in the sub-period of the 1997 Korean foreign currency crisis is greater than in all the other periods in **Table 3**, regardless of stock group. This is evidence suggesting that market crashes have a significant influence on the fat tails of stock return distributions. However, the fat tails cannot be fully explained by market crash since the sub-periods before and after the crash also have statistical probabilities greater than 0.5%, and the estimated parameter of Student's *t* distribution is smaller than the expected value of the normal distribution. In terms of stock groups, regardless of the period, the return distribution in the small-cap group has, on average, much fatter tails than the large-cap group. The results from the estimated parameter of Student's *t* distribution are the same as in Panel A. In addition, the positive tail in the return distribution is fatter, on average, than the negative tail in the Korean stock market. This result means that the large-scale price fluctuations that appear in the positive tail of the return distributions occur more frequently in the market, even though the large losses associated with market crashes appear in the negative tail of the distribution.

**Table 4** and **Figure 3** present the effect of volatility clustering on the fat tails in the return distributions using the GARCH(1,1) model as a filter model.[5] Using the standardized residuals estimated by the GARCH filter model, the fatness of the tails in the stock return distributions is empirically examined based on the same test procedure as that applied in **Table 3**. The fatness of the tails is again measured by the statistical probability and the estimated parameter of Student's *t* distribution. The results are separately presented by period and stock group. Panel A indicates the average values of the statistical probabilities related to the fatness of the tails in the standardized residual distribution for each stock group. Panel B presents the average values of the parameters of Student's t distribution that are directly estimated from the standardized residuals for each stock group. As in **Table 3**, the results are generated after removing the top 0.005 of stocks with the highest value and the bottom 5% of stocks having the lowest value among all the stocks selected in each period.

[Insert **Table 4** in here]

According to the results, regardless of period and stock group, the fatness of the tails in the standardized residual distribution produced in the GARCH(1,1) filter model is reduced, as compared to the results of **Table 3** using the original stock returns. For example, from **Table 3** to **Table 4** the statistical probability for the all-stocks group decreases from 1.30%→0.81% in the negative tail and from 1.69%→1.31% in the positive tail, and the estimated parameter of Student's *t* distribution increases from 2.098 → 3.254. These results are confirmed in **Figure 3**, where we use the error-bar plot to simultaneously compare both statistical probability and estimated parameter in the case of the all-stocks group, controlling for the different number of stocks in the periods. **Figure 3(a)** shows all statistical probabilities on the negative and positive tails observed in both **Tables 3** and **4**, and **Figure 3(b)** presents all estimated parameters of Student's *t* distribution in the two tables.

[Insert **Figure 3** in here]

**Figure 3** shows the reduction in the fatness of the tails in the return distribution using standardized residuals from the GARCH filter model, which reveals a decrease in the statistical probability and an increase in the estimated parameter, as compared to the fatness of the tails using the original stock returns. In addition, the reduction rate of the fatness of the tails is, on average, greater in the recent sub-periods than in the earlier sub-periods, and greater in the negative tail than in the positive tail. This suggests that volatility clustering is a significant factor influencing the fat tails of return distributions. However, these fat tails cannot be fully explained solely by volatility clustering since all the statistical probabilities remain greater than 0.005 after controlling for this factor. Furthermore, all estimated parameters of Student's $t$ distribution remain smaller than the expected value of the normal distribution. On the other hand, the 2007-2009 U.S. subprime mortgage crisis that was directly caused by the Lehman Brothers bankruptcy filing on September 15, 2008, triggered the collapse of the global financial market. All Asian financial markets, including Korea's, were influenced by this market crash (see, Wang [38]). Therefore, for robustness, we additionally verify the effect of the 2007-2009 U.S. subprime mortgage crisis on the results from the second sub-period (2006.7~2016.6, Sub2-P3). That is, the second sub-period is separated into two sub-periods: a sub-period of the U.S. subprime mortgage crisis (2007.6~2009.7, Sub3-P1) and a sub-period of the more recent normal market (2009.7~2015.6, Sub3-P2). We present the results on the far-right side of **Figure 3** with a large box. The results present evidence to definitely support the previous results.

---

[5] The previous studies employ various types of GARCH family model. For example, Stoyanov et al. [12] report the results using the ARMA(1,1)-GARCH(1,1) model as a filter model. In addition, the EGARCH model is utilized as an improved GARCH model. The EGARCH model suggested by Nelson [34] reflects the asymmetric behavior of volatility from bad news and good news, that is, leverage effect. To obtain robustness on the results in **Table 4**, we test the same procedure using two-type GARCH models: the best-fit ARMA(a,b)-GARCH(p,q) model for a=1,2, b=1,2 p=1,2, and q=1,2; and the best-fit EGARCH(p,q) model for p=1,2,3, and q=1,2,3. The AIC (Akaike information criteria) is utilized as a criterion to determine the best-fit model [35]. The result is reported in the Table of the Appendix. In the table, we definitely verify that the results are qualitatively the same. Accordingly, for the sake of parsimonious criterion and simplicity, we concentrate on reporting results from the standard GARCH(1,1) model in the paper.

## 4. SUMMARY AND CONCLUSIONS

This study used simple statistical methods to re-examine the existence of fat tails in stock return distributions in the Korean stock market. It controlled for market crashes and volatility clustering, and compared the empirical distribution of actual returns to various theoretical distributions. The results are summarised as follows. The fatness of the tails in stocks return distributions is greater in the recent sub-periods than in the pre-crash sub-periods. Fatness is also greater in the small-cap group than in the large-cap group. Although market crash and volatility clustering are crucial factors influencing the fat tails of return distributions, they cannot fully explain the existence of the fat tails since they remain after controlling for the influence of these two contributing factors. Our findings are robust regardless of the type of period or stock group. On the other hand, the fat tails of the return distribution might indicate economically that the many returns of extreme losses and extreme profits that deviated from the average value of returns are located in tail parts compared to the theoretical normal distribution. These extreme values lead to an increase in the magnitude of volatility, in particular, measured by the standard deviation of return. In the field of finance, volatility as a risk measurement plays an important role in portfolio optimal allocations and asset pricing models. Consequently, the existence of fat tails in the return distribution is strongly related to volatility. Accordingly, our findings may inspire future studies aimed at uncovering the properties of systematic and unsystematic risks encoded in the fat tails of stock return distributions, thereby acting as a bridge to connect the existing studies on tail risk to common factors in pricing models.

**ACKNOWLEDGMENTS**

This work was supported by (E. Scalas) JSPS Invitational Fellowship S18099 and JSPS KAKENHI 17K01270.

**Appendix. Table**.

. Existence of fat tails in the distribution of standardized residuals

|  | Entire period | Sub-period Type 1 | | Sub-period Type 2 | | |
|---|---|---|---|---|---|---|
|  | 1980.1~2015.6 | 1982.7~1997.6 | 2000.7~2015.6 | 1988.7~1997.6 | 1997.7~2006.6 | 2006.7~2015.6 |
| **Panel A: using the best-fitted ARMA(a,b)-GARCH(p,q) model (a, b = 1, 2; p, q = 1, 2)** | | | | | | |
| **All stocks** | | | | | | |
| negative tail | 0.008077 | 0.007801 | 0.008497 | 0.006768 | 0.008544 | 0.008690 |
| positive tail | 0.012960 | 0.011157 | 0.015514 | 0.009059 | 0.015078 | 0.014924 |
| **Large cap** | | | | | | |
| negative tail | 0.008204 | 0.007950 | 0.008376 | 0.007007 | 0.008465 | 0.008546 |
| positive tail | 0.012630 | 0.011478 | 0.013966 | 0.010099 | 0.014268 | 0.012849 |
| **Small cap** | | | | | | |
| negative tail | 0.007895 | 0.007489 | 0.008507 | 0.006618 | 0.008421 | 0.008724 |
| positive tail | 0.013518 | 0.010448 | 0.016912 | 0.008388 | 0.015675 | 0.016600 |
| **Panel B: using the best-fitted EGARCH(p,q) model (p, q = 1, 2, 3)** | | | | | | |
| **All stocks** | | | | | | |
| negative tail | 0.008739 | 0.008427 | 0.008659 | 0.006922 | 0.008578 | 0.008852 |
| positive tail | 0.013331 | 0.011529 | 0.015590 | 0.008868 | 0.014916 | 0.015039 |
| **Large cap** | | | | | | |
| negative tail | 0.008909 | 0.008439 | 0.008497 | 0.007187 | 0.008406 | 0.008614 |
| positive tail | 0.012906 | 0.011615 | 0.014051 | 0.009935 | 0.014078 | 0.013076 |
| **Small cap** | | | | | | |
| negative tail | 0.008658 | 0.008218 | 0.008713 | 0.006965 | 0.008564 | 0.008988 |
| positive tail | 0.013900 | 0.010891 | 0.016837 | 0.008115 | 0.015618 | 0.016682 |

Note: Table shows the results on the existence of fat tails in the distribution using standard residuals estimated from two-type GARCH models. The AIC criterion is employed to determine the best-fit model.

**Table 1**.

Summary statistic of stock returns

|  | **Entire period** | **Sub-period Type 1** |  | **Sub-period Type 2** |  |  |
|---|---|---|---|---|---|---|
|  | 1980.1~2015.6 | 1982.7~1997.6 | 2000.7~2015.6 | 1988.7~1997.6 | 1997.7~2006.6 | 2006.7~2015.6 |
| # of obs. | 8,719 | 3,661 | 3,710 | 2,196 | 2,216 | 2,231 |
| **All stocks** | | | | | | |
| # of stocks | 183 | 188 | 497 | 251 | 460 | 582 |
| ave. (x100) | 0.0247 | 0.0476 | 0.0307 | 0.0146 | -0.0140 | 0.0300 |
| st.dev. | 0.0319 | 0.0251 | 0.0320 | 0.0257 | 0.0428 | 0.0291 |
| skewness | 0.0011 | -0.4053 | 0.2549 | 0.0734 | 0.1556 | 0.2396 |
| kurtosis | 15.6705 | 21.9210 | 8.2482 | 5.3032 | 5.8426 | 8.8157 |
| **Large-cap** | | | | | | |
| # of stocks | 73 | 54 | 198 | 70 | 153 | 232 |
| ave. (x100) | 0.0348 | 0.0363 | 0.0417 | -0.0014 | 0.0014 | 0.0336 |
| st.dev. | 0.0290 | 0.0253 | 0.0290 | 0.0248 | 0.0420 | 0.0263 |
| skewness | -0.1003 | -0.3850 | 0.2009 | 0.1327 | 0.1851 | 0.1199 |
| kurtosis | 12.4881 | 16.0163 | 8.1587 | 5.0002 | 5.7568 | 8.4432 |
| **Small-cap** | | | | | | |
| # of stocks | 73 | 54 | 198 | 70 | 153 | 232 |
| ave. (x100) | 0.0173 | 0.0534 | 0.0180 | 0.0253 | -0.0274 | 0.0251 |
| st.dev. | 0.0340 | 0.0253 | 0.0344 | 0.0264 | 0.0437 | 0.0320 |
| skewness | -0.0453 | -0.6479 | 0.2989 | 0.0307 | 0.1500 | 0.3429 |
| kurtosis | 13.0036 | 21.1116 | 8.2849 | 5.5425 | 5.7926 | 9.0496 |

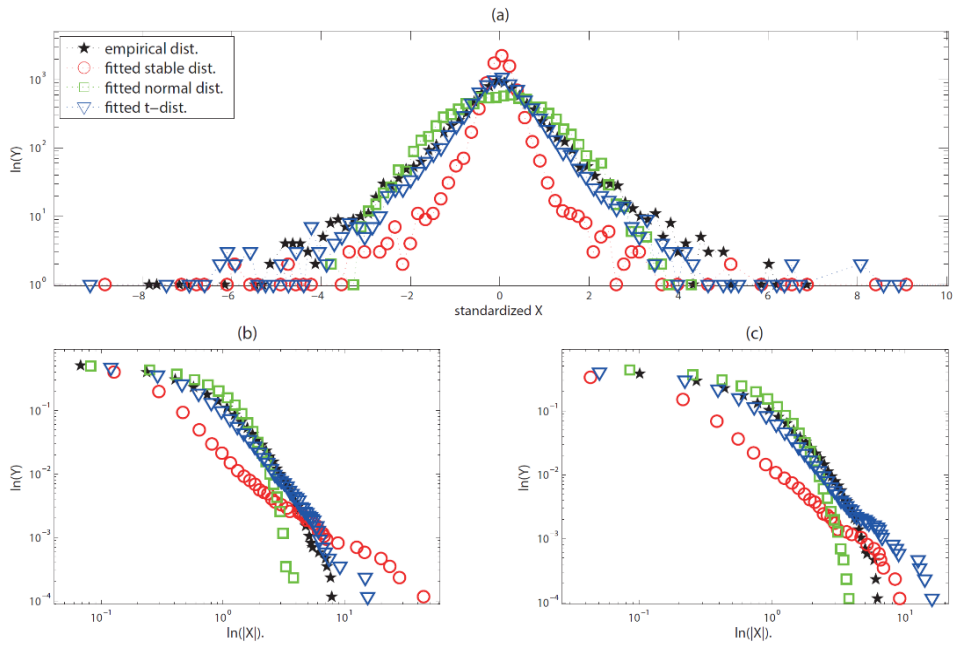

**Fig 1.** Comparison of three theoretical distributions and the empirical distribution. The figure compares three theoretical distributions and an empirical distribution (*) using market returns for KOSPI over the period from January 1980 to June 2015. The theoretical distributions in the figure are the normal distribution (□), α-stable distribution (○), and Student's *t* distribution (▽). In the figures, the x-axis of figure(a) indicates the standardized returns and the y-axis the logarithmic value of statistical probability; both the x- and y-axes in figures (b) and (c) indicate logarithmic values.

**Table 2.**

Estimated parameters of the theoretical distributions and Kolmogorov-Smirnov test

| estimated parameters | | | | Kolmogorov-Smirnov test | | | |
|---|---|---|---|---|---|---|---|
| | | | | statistic | p<0.01 | p<0.05 | p<0.10 |
| **Panel A: Normal distribution** | | | | | | | |
| ( μ ) | ( σ ) | | | | | | |
| 0.000339 | 0.016190 | | | 0.0813 | 100 | 100 | 100 |
| **Panel B: Student's t distribution** | | | | | | | |
| ( υ ) | ( μ ) | ( σ ) | | | | | |
| 2.75 | 0.000394 | 0.009835 | | 0.0325 | 67 | 85 | 87 |
| **Panel C: Stable distribution** | | | | | | | |
| ( α ) | ( β ) | ( γ ) | ( δ ) | | | | |
| 1.526500 | -0.013621 | 0.008022 | 0.000337 | 0.01943 | 100 | 100 | 100 |

Note: The table presents the following key parameters of the theoretical distributions. The normal distribution consists of parameters of the mean (μ) and standard deviation (σ). Student's t distribution has a parameter of the degrees of freedom (υ). Stable distribution consists of four parameters: the tail (α) and skewness (β) for the shape in the distribution, the scale (γ) and location (δ) in the distribution.

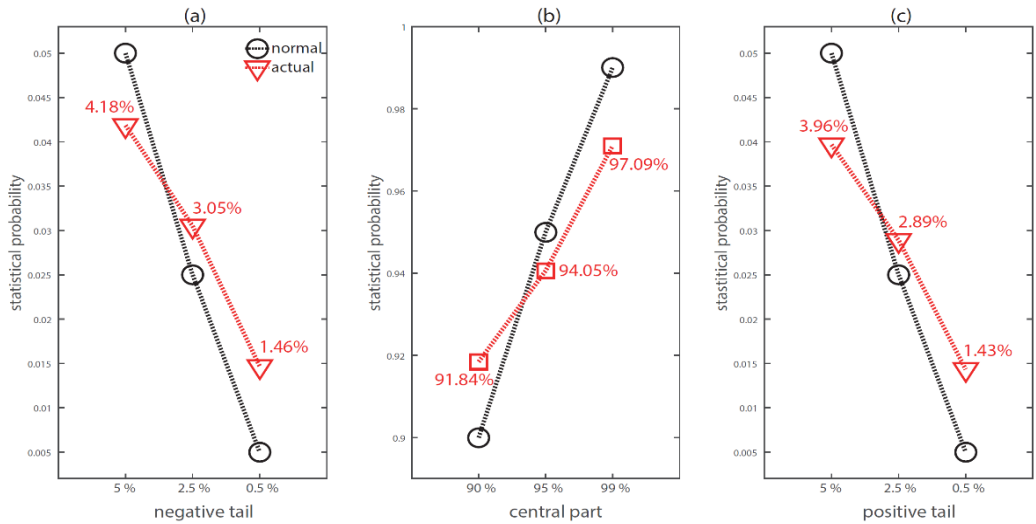

**Fig 2**. Comparison of statistical probabilities in the empirical and standard normal distributions. The figures show the statistical probability of the empirical and standard normal distributions in the central and tail parts based on **Figure 1**. The circles (◯) indicate statistical probabilities from the standard normal distribution, and the triangles (▽) and boxes (□) are statistical probabilities from the empirical distribution in the negative/positive tail and central parts, respectively. In Figure 2(b), the x-axis separately indicates the three-type statistical probability of 90%, 95%, and 99%. In Figure 2(a) and 2(c), the x-axis is divided by three-type statistical probability of 5%, 2.5%, and 0.5% for the negative and positive tails, respectively. The numbers on the circles, triangles and boxes in the figures present the statistical probabilities that are observed from market returns of KOSPI in the Korean stock market over period from January 1980 to June 2015.

**Table 3**.

Existence of fat tails in the distribution of stock returns

| | Entire period | Sub-period Type 1 | | Sub-period Type 2 | | |
|---|---|---|---|---|---|---|
| | 1980.1~2015.6 | 1982.7~1997.6 | 2000.7~2015.6 | 1988.7~1997.6 | 1997.7~2006.6 | 2006.7~2015.6 |
| **Panel A: Statistical probabilities, $f_N/f_T$, of the tails in the distribution** | | | | | | |
| **All stocks** | | | | | | |
| # of stocks | 165 | 170 | 443 | 225 | 413 | 523 |
| negative tail | 0.013035 | 0.008731 | 0.012818 | 0.008187 | 0.014475 | 0.012189 |
| positive tail | 0.016918 | 0.012330 | 0.019402 | 0.010524 | 0.019495 | 0.017438 |
| **Large-cap** | | | | | | |
| # of stocks | 64 | 48 | 178 | 62 | 137 | 207 |
| negative tail | 0.012642 | 0.008479 | 0.011920 | 0.007844 | 0.013353 | 0.011585 |
| positive tail | 0.015769 | 0.012212 | 0.016500 | 0.011032 | 0.017685 | 0.014575 |
| **Small-cap** | | | | | | |
| # of stocks | 65 | 48 | 178 | 62 | 137 | 208 |
| negative tail | 0.013187 | 0.008570 | 0.013760 | 0.008373 | 0.015093 | 0.012986 |
| positive tail | 0.018080 | 0.011825 | 0.022349 | 0.010092 | 0.020752 | 0.020358 |
| **Panel B: Estimated parameter of Student's t distribution (degree of freedom)** | | | | | | |
| All stocks | 2.0985 | 2.8359 | 2.4851 | 6.8115 | 2.5223 | 2.6168 |
| Large-cap | 2.1733 | 2.8778 | 2.9345 | 5.9157 | 2.8827 | 3.1463 |
| Small-cap | 1.9982 | 2.7933 | 2.1618 | 7.3395 | 2.3287 | 2.2179 |

**Table 4**.

Existence of fat tails in the distribution of standardized residuals by GARCH(1,1)

|  | Entire period | Sub-period Type 1 | | Sub-period Type 2 | | |
|---|---|---|---|---|---|---|
|  | 1980.1~2015.6 | 1982.7~1997.6 | 2000.7~2015.6 | 1988.7~1997.6 | 1997.7~2006.6 | 2006.7~2015.6 |
| **Panel A: Statistical probabilities, $f_N/f_T$, of the tails in the distribution** | | | | | | |
| **All stocks** | | | | | | |
| # of stocks | 164 | 169 | 445 | 223 | 414 | 524 |
| negative tail | 0.008117 | 0.007786 | 0.008515 | 0.006863 | 0.008544 | 0.008797 |
| positive tail | 0.013109 | 0.011075 | 0.015609 | 0.009075 | 0.015128 | 0.014934 |
| **Large cap** | | | | | | |
| # of stocks | 65 | 48 | 178 | 61 | 136 | 207 |
| negative tail | 0.008315 | 0.007864 | 0.008456 | 0.007124 | 0.008537 | 0.008653 |
| positive tail | 0.012811 | 0.011290 | 0.013995 | 0.010216 | 0.014259 | 0.012853 |
| **Small cap** | | | | | | |
| # of stocks | 64 | 48 | 178 | 61 | 136 | 208 |
| negative tail | 0.007948 | 0.007512 | 0.008501 | 0.006779 | 0.008355 | 0.008813 |
| positive tail | 0.013604 | 0.010351 | 0.016930 | 0.008432 | 0.015713 | 0.016615 |
| **Panel B: Estimated parameter of Student's t distribution (degree of freedom)** | | | | | | |
| All stocks | 3.2538 | 3.6202 | 3.5489 | 10.0008 | 4.0429 | 3.7753 |
| Large cap | 3.3252 | 3.6080 | 4.2668 | 8.4240 | 4.6040 | 4.8184 |
| Small cap | 3.1236 | 3.6847 | 3.0637 | 11.9901 | 3.7589 | 3.0734 |

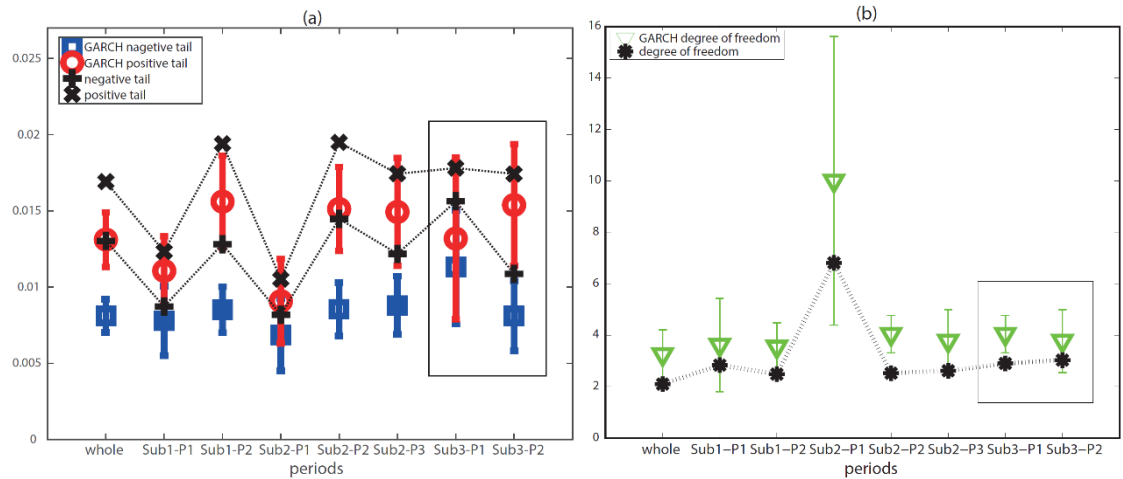

**Fig. 3.** Comparison of the fatness of stock returns and standardized residuals using GARCH (1,1). The figure shows the fatness of the negative and positive tails of the distribution observed in **Table 3** using actual stock returns and in **Table 4** using the standardized residual data estimated by the GARCH (1,1) filter model. The results are for the case in which stocks belonging to the all-stocks group are used. The error-bar plot controls for the difference in the number of stocks in each period. Figure (a) compares the results of **Table 3** (+, x) and the results of **Table 4** (○, □) based on the statistical probability. Figure (b) presents the comparison based on the estimated parameters of Student's t distribution observed from the results of **Tables 3** (*) and **4** (▽). The large boxes on the right side of the figures show the results of two sub-periods separated from the second sub-period (2006.7~2016.6, Sub2-P3); that is, the U.S. subprime mortgage crisis (2007.7~2009.6, Sub3-P1) and after the market crisis (2009.7~2015.6, Sub3-P2).